\documentclass[aps,prl,preprint,groupedaddress,showpacs,nobibnotes,tightenlines]{revtex4}
\usepackage{latexsym,amsmath}

\begin{document}

\preprint{OKHEP-01-08}

\title{Bulk versus brane running couplings}

\author{Kimball A. Milton}
\email[]{milton@mail.nhn.ou.edu}
\affiliation{Department of Physics and Astronomy, The University of 
Oklahoma, Norman 73019 USA}

\author{Sergei D.  Odintsov}
\email[]{odintsov@ifug5.ugto.mx} 
\altaffiliation{Tomsk State Pedagogical University, 
Tomsk, RUSSIA}
\affiliation{Instituto de Fisica de la Universidad de Guanajuato, 
Lomas del Bosque, Leon, MEXICO}

\author{Sergio Zerbini}
\email[]{zerbini@science.unitn.it}
\affiliation{Department of  Physics and Gruppo 
Collegato INFN\\
University of Trento, ITALY}

\date{\today}

\begin{abstract}
A simplified higher dimensional Randall-Sundrum-like model in 6 
dimensions is 
considered. It has been observed previously by Goldberger and Wise that
 in such a self-interacting scalar theory on the bulk with a conical 
singularity there is mixing of renormalization of 4d brane couplings 
 with that of the bulk couplings.
We study the influence of the running bulk couplings on the running of the
4d brane couplings.  We find that bulk quantum effects may 
completely alter the running of brane couplings. In particular, the 
structure of the Landau 
pole may be drastically altered and  non-asymptotically free running may turn 
into asymptotically safe (or free) behavior.
\end{abstract}


\maketitle

\section{Introduction}
Recently Randall-Sundrum brane-world models \cite{RS} have attracted a great 
deal of interest in particle physics and phenomenology. One reason is 
that such models may resolve the hierarchy problem in a quite natural way.
Moreover, they may lead to some interesting changes in the
phenomenology of the Standard Model and related theories. 

A typical modern approach to quantum high energy theory is to consider
quantum fields on a
higher dimensional manifold (the bulk) in the presence 
of extended defects (the boundaries). For example, we could consider a 
3-brane localized in a 4-dimensional submanifold. On the bulk manifold, 
as  well as on the brane, there
exist divergences which result in the running of coupling constants 
in the standard way.
However, it has been known for some time that for spacetimes with boundaries 
there are not only the usual volume coupling constants but also
there are surface coupling constants \cite{surface}.
Of course, these influence each other; for example,  volume interactions
are reflected in surface terms, etc \cite{tsoupros}.
However, so far no drastic consequences have been discovered.
Nevertheless, it could be that the situation is different in some brane-world 
models. For example, one may 
wonder how the running of the bulk couplings influences the running on the 
brane and vice versa. With regard to this issue, a very simple 6d model
with a 4d brane has been proposed in Ref.~\cite{wise}. However, only
renormalization of bulk and brane coupling constants has been 
investigated there.
Here we reconsider this model, making use of  heat-kernel techniques
and perform estimates of the running of bulk and brane couplings.
It will be seen that the bulk running couplings can significantly influence 
the running of the 4d cosmological constant, 4d mass and 4d self-scalar 
interaction. In particular, instead of a constant behavior of 
the cosmological 
constant (in the sense of the renormalization group), we find a running,
such that the cosmological constant increases with energy.
The Landau pole for the brane mass and the scalar self-interaction
is drastically altered by the bulk running couplings, and in general
becomes a cut.

\section{The Model}

Our toy model consists of a massive Euclidean self-interacting scalar in 
a 6-dimensional space with a conical singularity, due to the
presence of 3-brane. The metric is chosen to be
\begin{equation}
ds_6^2=dr^2+r^2d\theta^2+ds^2_0\,,
\label{111}
\end{equation}
where $ds^2_0$ is the 4-dimensional flat metric,  
the brane is located at $r=0$, and $\theta$ has a period $\beta$, $\beta $ being
the deficit angle of the cone. When $\beta=2 \pi/N$, $N$ a positive integer,
one is dealing with a less singular manifold, namely an orbifold, 
while for $N=1$, $\beta= 2\pi$, one  has the smooth 2-dimensional plane.
The action  reads
\begin{equation}
S= \int d^6x\, \sqrt{g} \left[ \phi  \frac{1}{2}(-\Box_6+m^2  ) \phi+
V(\phi) \right]+\int d^4x\, W(\phi)\,,  
\label{2}
\end{equation}
where $V(\phi)=\frac{g_4}{4!}\phi^4+\dots$ denotes a series of 
scalar  bulk couplings. We also introduce 
 a ``surface" term which depends on surface scalar couplings
\begin{equation}
W(\phi)=\left[\lambda_0+\frac{\lambda_2}{2} \phi^2+\frac{\lambda_4}{4!} 
\phi^4+\dots\right]\,,  
\end{equation}
namely it may contain a brane tension $\lambda_0$, a brane mass  
$\lambda_2$, a $\phi^4$ coupling $\lambda_4$, as well as higher terms.
As we will see, these surface terms are necessary because we are dealing 
with a manifold with a conical singularity (see also Ref.~\cite{georgi}).
We also assume that the brane is not dynamical, namely we are dealing 
with a rigid brane, and therefore we  neglect the brane kinetic term.

The total one-loop action operator consists of a bulk part and a contact 
delta-function singular potential. First, we discuss briefly the problems 
associated with 
this latter contact term. For a nice discussion from a diagrammatic point of 
view, see the approach contained in Ref.~\cite{wise}.
Here one may follow the standard and well-known   argument used in 
nonrelativistic 
quantum mechanics and put on a rigorous basis in 
Refs.~\cite{faddeev,albeverio} (see also references cited therein).

Briefly, the issue to be addressed is based on the following remarks. 
When one has  to evaluate the Green's functions (i.e. two-point functions or 
correlation functions)  related to a Laplace operator ``perturbed" by a 
delta-function potential, typically one is forced to deal with the product of 
a delta-function distribution and a two point-function, 
which can have singularities on 
the support of the delta function. Only in codimension 1 (RS-like models) 
is this singularity absent, since then 
the two-point function is not singular at 
the coincidence limit of the two points.  

The method to be used \cite{faddeev} consists in the 
regularization of the delta function by means of a  sequence of smooth  
functions depending on 
a cutoff parameter. The key point is to assume that the bare coupling
constants (which can be thought of as strengths of the terms in
the delta-function
potential) also depend on the regularization parameter.  
Then, one can easily solve  the 
functional equation for the two-point function within this regularization. 
In order to remove the 
cutoff parameter dependence, one has to assume that the contact divergence, 
present if the 
codimension is bigger than one, is cured  by a suitable compensating 
behavior of the coupling. If the
codimension is two, the divergence is only logarithmic and, 
for dimensional reasons,  there is room for the appearance  of an 
arbitrary renormalization parameter $\mu$. 
In general, this happens when the codimension is even.  
As a result, the renormalized coupling constants will depend on $\mu$ and 
it is not difficult to demonstrate that they must satisfy renormalization 
group equations.

\section{One-loop correction and renormalization group}
 
The one-loop correction is determined by the  total one-loop fluctuation 
operator, which  reads 
\begin{equation}
L_6 = -\Box_\beta -\Box_4+M^2+W''(\Phi)\delta^{(2)}(x)\,, 
\label{3}
\end{equation}
where $\Box_\beta$ is the 2-dimensional Laplacian on the cone, $ \Phi $ 
is the background field and $M^2=m^2+V''(\Phi)$ is an 
effective mass.

We shall make use of zeta-function regularization and related heat-kernel 
techniques (see, for example Refs.~\cite{elib,bytsenko96}).
Within the one-loop approximation,  we have to evaluate the
zeta-function at zero, namely $\zeta(0|L_6)$, since this quantity 
gives rise to the one-loop divergences and governs the one-loop beta functions.
There are also contributions due to the conical singularity and the brane 
delta-function contribution, which gives additive 
contributions to  $\zeta(0|L_6)$, which have been diagrammatically 
evaluated in Ref.~\cite{wise}.

The zeta-function for the operator $L_6$ is defined by 
\begin{equation}
\zeta(s|L_6)=\frac{1}{\Gamma(s)}\int_0^\infty dt\, t^{s-1}\mbox{Tr}e^{-tL_6}\,.
\end{equation}
In the presence of the conical singularity and delta-function potential,
one has for small $t$
\begin{eqnarray}
\mbox{Tr}e^{-tL_6}&=&\int d^6x \frac{e^{-tM^2}}{(4\pi)^3t^3}
\left[ 1+b_2(x)t^2+b_3(x)t^3+ O(t^4)\right]\nonumber\\
&&\quad\mbox{}+\int d^4x \frac{e^{-tM^2}}{(4\pi)^2t^2}(I_\beta+I_W)
\left[ 1+b_2(x)t^2+b_3(x)t^3+ O(t^4)\right]
\,,
\end{eqnarray}
where $b_2(x)$ and $b_3(x)$ depend only on the derivative of the 
effective mass and are given by
\begin{subequations}
\begin{eqnarray}
b_2(x)&=&-\frac{1}{6}\Box M^2\,,\,\,\\
b_3(x)&=&-\frac{1}{60}\Box^2 M^2+\frac{1}{12}\partial_\mu M^2
\partial^\mu M^2\,.
\end{eqnarray}
\end{subequations}

We also have included the boundary terms due to the conical singularity and 
the delta-function contact term. The first  depends on the numerical 
quantity \cite{cone} 
\begin{equation}
I_\beta=\frac{1}{12}\left( \frac{2\pi}{\beta}-\frac{\beta}{2\pi}\right)
\end{equation}
Of course, this boundary term is  absent if $\beta=2\pi$, $I_\beta=0$, 
 namely in the absence of the conical singularity.

 The contact contribution depends on the quantity $W''$, and in the 
perturbative regime, corresponding to $W''$ ``small,'' is, 
as a function of the background field, simply given by
\begin{equation}
W''(\Phi)=\left[\lambda_2 \Phi^2+\frac{1}{2}\lambda_4 \Phi^4+
\frac{1}{4!}\lambda_6\Phi^6+....
 \right]\,,  
\label{con}
\end{equation}

The analytic continuation of the zeta function to $s=0$ is
\begin{eqnarray}
\zeta(0|L_6)&=&\int d^6x \frac{1}{(4\pi)^3}  
\left[-\frac{(M^2)^3}{6}-b_2(x)M^2+b_3(x)
\right]\nonumber\\
&&\quad\mbox{}+\int
 d^4x \frac{1}{(4\pi)^2}(I_\beta+I_W)\left[\frac{(M^2)^2}{2}+b_2(x)\right].
\end{eqnarray}
The heat kernel coefficients $a_n$
 are defined by the small-$t$ expansion of the
heat kernel, ($d$ is the dimension of the manifold, here $d=6$)
\begin{equation}
\mbox{Tr}e^{-tL_6}\sim{1\over(4\pi t)^{d/2}}\sum_na_nt^n,\quad t\to 0+,
\end{equation}
 so the integrand in the first term of the above expression is the $a_3(x)$ 
coefficient, which reads
\begin{equation}
a_{3,{\rm bulk}}(x)
=-\frac{(M^2)^3}{6}+\frac{M^2 \Box M^2}{6}+
\frac{\partial_\mu M^2\partial^\mu M^2}{12}-\frac{ \Box^2 M^2}{60}\,.
\label{a3}
\end{equation}
For the sake of completeness, we note that
the $a_2(x)$ coefficient, which is relevant in four dimensions, is
\begin{equation}
a_{2}(x)=\frac{(M^2)^2}{2}-\frac{\Box M^2}{6}\,.
\label{a2}
\end{equation}
These  expressions actually define the structure of the one-loop bulk 
and surface divergences.
It is already clear that the theory under consideration is not 
renormalizable  at the one-loop level! In such a situation, we will be 
interested in obtaining the one-loop renormalization for ``renormalizable''
bulk interactions which will appear also in 
the renormalization of brane couplings. These are the mass and 
four-scalar bulk couplings.
They are actually defined by the term in $a_{3,{\rm bulk}}$
cubic in the effective mass.
The corresponding one-loop renormalization group (RG) equations have the form
\begin{subequations}
\begin{eqnarray}
\frac{dm^2}{dt}&=&c_1 m^4 g_4\,,\\
\frac{ dg_4}{dt}&=&c_2 m^2 g_4^2\,,
\end{eqnarray}
\end{subequations}
where 
\begin{equation}
c_1=\frac{1}{2}\frac{1}{(4\pi)^3}\,,\,\,c_2=\frac{3}{(4\pi)^3}\,,
\end{equation}
and $t$  stands for $ \ln \frac{\mu}{\mu_0}$. 
The solutions of these equations are not difficult to find, since their 
product $p(t)=g_4(t)m^2(t)$ satisfies ($c=c_1+c_2$)
\begin{equation}
\frac{d p}{dt}=cp^2\,.
\end{equation}
As a result, ($p_0=p(0)$)
\begin{subequations}
\begin{eqnarray}
p(t)&=&\frac{p_0}{1-cp_0 t}\,,
\label{p}\\
m^2(t)&=&m^2(0)(1-cp_0 t)^{-c_1/c}\,,
\label{m}
\\
g_4(t)&=&g_4(0)(1-cp_0 t)^{-c_2/c} \,.
\label{g}
\end{eqnarray}
 \end{subequations}
 As  $p_0$ is positive, we have to consider the regime 
$1 >cp_0 t$. When the denominator in above equations becomes zero
we see the appearance 
of a Landau pole in $p(t)$ in terms of the initial mass and self-interaction.
(In $m^2$ and $g_4$ the singularity is a branch point.)
As one sees, the infrared (IR) ($t\to-\infty$) behavior 
of above RG couplings is asymptotically free (AF).

A more interesting scenario emerges if we recognize that 
our observable physics is defined on a 4d brane.
Whatever the physics is in six dimensions, we require that it
be sensible for a 4d-brane observer.
For example, let us imagine that we could take the wrong sign for 
the scalar self-interaction
in six dimensions. Then, the RG equation for $g_4$ appears with the 
opposite sign.
In other words, the RG solution for $g_4$ appears with a positive
 sign in the denominator
 and becomes asymptotically free in the UV. The same occurs with the RG mass.
 Thus, we may search for consequences on the brane
of both AF and non-AF RG solutions.

Proceeding onwards, we see that the regularized one-loop effective action reads
\begin{equation}
\ln \det \frac{L_6}{\mu^2}(\varepsilon)=-\frac{\mu^{2\varepsilon}}{2}
\int_0^\infty dt\,t^{\varepsilon-1}
\mbox{Tr}e^{-tL_6}=- \frac{\mu^{2\varepsilon}}{2}
\Gamma(\varepsilon)\zeta(\varepsilon|L_6)\,,
\end{equation}
where an arbitrary mass scale $\mu$ has been introduced to keep each expression
dimensionless.
Thus, for small $\varepsilon$, 
\begin{equation}
\ln \det \frac{L_6}{\mu^2}(\varepsilon)=- \frac{1}{2}\left[
\frac{\zeta(0|L_6)}{\varepsilon}+(\ln \mu^2+\gamma)\zeta(0|L_6)+\,
\zeta'(0|L_6)+{\cal O}(\varepsilon)\right].
\end{equation}

One can see that due to the presence of the conical singularity, there are 
``brane"  surface  and contact contributions in the divergence as well as in 
the finite part 
depending on the scale $\mu^2$. These  surface contributions have to be 
added to the bulk counterterm in order to remove the additional ultraviolet 
divergences related to the conical singularity and contact term.  
These additional terms
also modify the one-loop renormalization group equations  and depend on 
the bulk potential $V(\phi)$ and the brane potential $W(\phi)$.

Let us now write the RG equations for the surface (brane) couplings which were 
derived in Ref.~\cite{wise}. For the special choice of the deficit angle 
$\beta=\pi$ (a $Z_2$ orbifold),  they read at one-loop,
\begin{subequations}
\begin{eqnarray}
\frac{d \lambda_0}{d t}&=&\frac{m^4}{256\pi^2}-\frac{m^4\lambda_2}{64\pi^3}\,, 
\label{22}\\   
\frac{d \lambda_2}{d t}&=&
\frac{\lambda_2^2}{\pi}+
\frac{m^2 g_4}{128\pi^2}-\frac{m^4\lambda_4}{64\pi^3}\,.
\label{33}
\end{eqnarray}
We also include the renormalization group
equation for the coupling constant related to the 
brane four-point function, the last term of which was not given 
explicitly in Ref.~\cite{wise}:
\begin{equation}
\frac{d \lambda_4}{d t}=\frac{4\lambda_2\lambda_4}{\pi}+
\frac{3 g_4^2}{128\pi^2}-\frac{m^4\lambda_6}{64\pi^3}\,. 
\label{24}
\end{equation}
\end{subequations}   
In the above equations the last terms have their origin in the delta-function 
contribution (\ref{con}).

First, let us find the solutions of these RG equations when the one-loop 
corrections are not included, i.e., we drop the bulk terms involving $m$
and $g_4$. 
Note that  $\lambda_0$ is then a constant.  
The other two couplings  are:
\begin{subequations}
\begin{eqnarray}
\lambda_2(t)&=&\frac{\lambda_2(0)}{1-\frac{\lambda_2(0)}{\pi} t}\,.
\label{02}\\
\lambda_4(t)&=&\lambda_4(0)\left(1-{\lambda_2(0)t\over\pi} \right)^{-4}\,.
\label{03}
\end{eqnarray}
\end{subequations}
As one can see, we find non-asymptotically free behavior 
for these coupling constants. At high energies there is a
Landau pole defined by the initial RG mass.
At the same time, as usual, in the IR limit we have AF behavior for both 
coupling constants.
 
In order to investigate the role of the one-loop quantum corrections, let us 
consider only the surface terms and neglect the contact terms.
That is, we consider the truncated equations (with $p=m^2g_4$) 
\begin{subequations}
\begin{eqnarray}
\frac{d \lambda_0}{d t}&=&\frac{m^4}{256\pi^2}\,, 
\label{223}
\\
\frac{d \lambda_2}{d t}-\frac{\lambda_2^2}{\pi}&=&
\frac{p}{128\pi^2}\,,
\label{333}
\\
\frac{d \lambda_4}{d t}-\frac{4\lambda_2\lambda_4}{\pi}&=&
\frac{3 g_4^2}{128\pi^2}\,. 
\label{244}
\end{eqnarray}
\end{subequations}
The first of these equations can be easily integrated,
\begin{equation}
\lambda_0(t)- \lambda_0(0)=\frac{ m^4(0)}{256 \pi^2}
\frac{1}{(c_2-c_1)p_0}\left[1- \left(1-cp_0 t\right)
^{\frac{c_2-c_1}{c}}\right]\,.
\label{ee}
\end{equation}
This reveals a very interesting running behavior 
for the effective cosmological constant. In the absence 
of the bulk contribution the cosmological constant
was unchanged from its initial RG value.  However,
after taking into account the bulk running, the evolution of
the cosmological constant is completely altered.
As long as $(c_2-c_1)/c$ is not an integer, which is the case here,
this solution still possesses a branch point at a suitably large value
of $t$; moreover,  it is not asymptotically free in either
the IR nor in the UV, but for small $t$ grows linearly with $t$.
Because $c_2>c_1$, it becomes large in both the IR and UV limits.

The third equation can be integrated once a 
solution of the second is known.
Although the latter is a nonlinear inhomogeneous differential equation 
belonging to the 
Riccati class, the general solution of which is not known, 
in our specific case, one can find an explicit and exact solution
(see Appendix).

Let us first  consider two limits.  
For small $t$ we can develop a perturbative (power series) solution in $t$,
\begin{equation}
\lambda_2(t)=\lambda_2(0)+\sum_{n=1}^\infty d_nt^n,
\label{pert}
\end{equation}
where from Eqs. (\ref{333}) and (\ref{p}) we easily find
\begin{subequations}
\begin{eqnarray}
d_1&=&b+{\lambda_2(0)^2\over\pi},\\
d_2&=&{\alpha b^2\over2\pi}+{b\over\pi}\lambda_2(0)+{1\over\pi^2}
\lambda_2(0)^3,\\
d_3&=&{\alpha^2b^3\over3\pi^2}+{b^2\over3\pi}+{\alpha b^2\over3\pi^2}
\lambda_2(0)+{4b\over3\pi^2}\lambda_2(0)^2+{\lambda_2(0)^4\over\pi^3},
\end{eqnarray}
\end{subequations}
and so on.  Here we have defined
\begin{equation}
b={p_0\over128\pi^2},\quad \alpha={\pi\over b}p_0 c,
\end{equation}
where in our case $\alpha=7$.

The other interesting limit is large $t$, that is $t\to-\infty$ if $p_0>0$
or $t\to\infty$ if $p_0<0$.  Then because 
\begin{equation}
p(t)\sim-{1\over ct},\quad p_0t\to-\infty,
\end{equation}
we find by achieving the only balance possible in Eq.~(\ref{333}) that
\begin{equation}
\lambda_2(t)\sim\pm\sqrt{|b|\pi\over1+|b|\alpha t/\pi},\quad
p_0<0,\,\, t\to\infty.
\label{lambda2sqrt}
\end{equation}
As one sees the Landau pole found in Eq.~(\ref{02}) 
 seems to disappear from our running behavior.
The mass is decreasing with the RG scale.

Finally, the solution of Eq.~(\ref{244}) reads
\begin{eqnarray}
\lambda_4(t)&=&\frac{3}{128\pi^2}e^{\frac{4}{\pi}\int_0^t dt' 
\lambda_2(t')}
\left[\int_0^t dt'\, e^{-\frac{4}{\pi}\int_0^{t'} dt_1 \lambda_2(t_1)}
 g_4^2(t') \right]\nonumber\\
 &&\quad\mbox{}+\lambda_4(0)e^{{4\over\pi}\int_0^t dt'\,\lambda_2(t')}.
\end{eqnarray}
We can easily insert the perturbative solution (\ref{pert}) 
into this expression to obtain the small $t$ behavior of $\lambda_4$.
The leading term is 
\begin{equation}
\lambda_4(t)=\lambda_4(0)+\left[{3\over128\pi^2}g_4(0)+{4\over\pi}
\lambda_2(0)\lambda_4(0)\right]t+{\cal O}(t^2).
\end{equation}
This, of course, immediately follows from Eq.~(\ref{244}).

More interesting again is the large $t$ behavior.  The leading behavior
can be readily determined from Eq.~(\ref{244}) by seeing, when the 
large $t$ behavior coming from Eq.~(\ref{g}) is inserted, which two terms
in the differential equation can balance.  It is clear that the only balance
possible is between the second and third terms, 
so we immediately find
\begin{equation}
\lambda_4(t)\sim{3\over32\sqrt{2\pi |p_0|}}g_4(0)^2(1-cp_0t)^{{1\over2}-
{2c_2\over c}},\quad p_0<0,\,\,t\to\infty.
\label{lambda4}
\end{equation}
Since ${1\over2}-{2c_2\over c}=-{5\over14}<0$,
this says that in the UV the four-scalar coupling constant 
 decreases. This is in the contrast with the non-AF behavior
seen in Eq.~(\ref{03}) without taking into account the bulk running!

These results are generalized by using
 the results of the analysis given in Appendix.
According to the general solution it follows that if we started from 
non-AF bulk running coupling (the case $p_0>0$)
then the structure of the Landau singularity for the brane 
mass (and here also for the brane four-coupling) is completely changed!
Previously in Eq.~(\ref{02}), 
it was defined by the initial RG value of the brane mass $\lambda_2(0)$. 
Now, there are an infinite number of singularities defined
by the initial RG values of the bulk mass and 
the bulk four-scalar coupling!  That is, there is a singularity (in general,
a branch point) at $t=t_0$, where
\begin{subequations}
\begin{equation}
t_0={1\over cp_0}={\pi\over\alpha b},
\end{equation}
and poles at $t=t_n$,
\begin{equation}
t_n=t_0-{\alpha\over4}z_n^2,
\end{equation}
\end{subequations}
where $z_n$ is the $n$th zero of a Bessel function.
Consequently, if one starts from the AF behavior for the bulk couplings 
(case $p_0<0$),
there are no singularites for $t>0$. That is, the Landau pole of the
brane couplings has been eaten by bulk quantum effects. The running behavior 
of the brane couplings becomes completely asymptotically safe!
Of course one could present corresponding numerical results for all couplings. 
We will not do this as the model under discussion is 
over-simplified and hardly realistic.

The important lesson which follows from our discussion is follows.
Even though the brane observer does not know much about the bulk wherein his 
brane is embedded, there exists a clear influence of bulk quantum physics on
the quantum physics on the brane. In particular, one may conjecture that 
there may be
the realistic (Standard Model-like) theories where observable running effects
 like asymptotic freedom are provided by bulk running couplings.
It is now a challenge to construct such brane-world models.

\begin{acknowledgments}
The research of KAM has been supported in part by a grant from the US
Department of Energy (\#DE-FG03-98ER41066), that of
 SDO  in part by 
CONACyT (CP, Ref.~990356) and in part by INFN, and that of SZ by INFN.
SZ thanks G. Cognola for helpful discussions.
\end{acknowledgments}

\appendix
\section{Solution of Riccati equation}

In this Appendix, we show how the Riccati equation (\ref{333}) can be solved.
According to the general theory of ordinary differential equations 
(see, for example, Ref.~\cite{ince}), it is 
known that  solutions of  
the Riccati equation can be found as soon as one is able to find solutions 
of a related  linear second order differential equation. In our case,
we may rewrite Eq.~(\ref{333}) in the form
\begin{equation}
\lambda_2'(x)+\frac{\lambda_2^2(x)}{\pi c p_0}+\frac{1}{128\pi^2c x}=0\,.
\end{equation}
with $'=\frac{d}{dx}$ and $x=1-cp_0t$. The associated linear 
second order equation simply  reads
\begin{equation}
\frac{d^2 u}{d^2 x}+
\frac{B u}{x}=0\,,
\label{3330}
\end{equation}
with  $B=\frac{\pi}{\alpha^2b}$.  The corresponding  
solution of the Riccati equation is  
\begin{equation}
\lambda_2(x)=\alpha b{d\over dx}\ln u(x).
\label{lambdainu}
\end{equation}
 Fortunately, Eq.~(\ref{3330})
can be reduced to a Bessel equation. In fact,  putting $z=2\sqrt{B x}$
and
\begin{equation}
u(x(z))=\frac{z}{2\sqrt{B }}Z_1(z)\,,
\label{Z}
\end{equation}
a simple computation leads to 
\begin{equation}
\frac{d^2 Z_1}{d^2z}+\frac{1}{z}\frac{d Z_1}{dz}
+\left(1-\frac{1}{z^2}\right)Z_1=0,
\label{bf}
\end{equation}
which is the Bessel equation of order $\nu=1$. 

The solution to Eq.~(\ref{bf}) is
\begin{equation}
Z_1(z)=J_1(z)+AN_1(z),
\end{equation}
where $J_1$ is the Bessel function of the first kind of order 1, and $N_1$
is the Neumann function of order 1.  The arbitrary constant of integration
is $A$.  $\lambda_2$ is now given by Eq.~(\ref{lambdainu}) with
\begin{equation}
u(x)=\sqrt{x}Z_1\left(2\sqrt{Bx}\right).
\end{equation}
The constant $A$ is determined by the initial value of $\lambda_2$:
\begin{equation}
\lambda_2(t=0)=\alpha b{d\over dx}\ln u(x)\bigg|_{x=1}.
\end{equation}

What are the singularities of $\lambda$?  If $A=0$, $Z_1$ has a zero at
$z=0$; otherwise it has a branch point there.  This means that there is
a singularity at $x=0$ or at
\begin{equation}
t_0={1\over cp_0}.
\end{equation}
This is the reflection of the Landau pole in $p(t)$, Eq.~(\ref{p}).
But there an an infinite number of other singularites of $\lambda_2$,
occurring where $Z_1$ possesses zeros; call these zeroes $z_n$.
(Asymptotically,
\begin{equation}
z_n=\arctan{1+A\over1-A}+n\pi,
\end{equation}
where the arctangent is the principal value and $n$ is an integer.)
The corresponding values of $t_n$ are
\begin{equation}
t_n=t_0-{\alpha\over4}z_n^2.
\end{equation}
Thus, if $p_0>0$, to the right of $t=0$ there are a finite number of
singularities, but an infinite number to the left; for $p_0<0$ all the 
singularities occur for $t<0$.  So let us examine the latter situation.
If $p_0<0$, $B<0$, so we write the solution in terms of modified Bessel
functions (Bessel functions of imaginary argument),
\begin{subequations}
\begin{eqnarray}
Z_1(iz)&=&I_1(z)+AK_1(z),\\
u(x)&=&\sqrt{x}\left[I_1\left(2\sqrt{|B|x}\right)+AK_1\left(2\sqrt{|B|x}\right)
\right].
\end{eqnarray}
\end{subequations}
Up to a constant factor, asymptotically,
\begin{equation}
u(x)\sim e^{2\sqrt{|B|x}}+Ae^{-2\sqrt{|B|x}},\quad x\to\infty.
\end{equation}
Thus we find the leading behavior
\begin{equation}
\lambda_2(t)\sim-\sqrt{\pi|b|\over1-cp_0t},\quad p_0<0,\,\,t\to\infty.
\end{equation}
This is consistent with the asymptotic result (\ref{lambda2sqrt}), 
and therefore also implies the behavior seen in Eq.~(\ref{lambda4}).



\end{document}